\documentclass[11pt]{article}
\usepackage{graphicx}
\begin{document}
\title{Can Barrier to Relative Sliding of Carbon Nanotube Walls Be Measured?}
\maketitle

\begin{center}
Andrey M. Popov\footnote{e-mail: popov-isan@mail.ru},  Yurii E. Lozovik

Institute of Spectroscopy, Fizicheskaya Str. 5, Troitsk, Moscow region, 142190, Russia

Evgenii K. Krivorotov

Moscow State Pedagogical University, Malaya Pirogovskaya Str. 1, Moscow, 119991, Russia
\end{center}

\begin{abstract}
Interwall interaction energies, as well as barriers to relative sliding of the walls along the nanotube axis, are first calculated for pairs of both
armchair or both zigzag adjacent walls of carbon nanotubes with a wide range of radiuses. It is found that for the pairs with the radius of the outer
wall greater than 5 nm both the interwall interaction energy and barriers to the relative sliding per one atom of the outer wall only slightly
depends on the wall radius. A wide set of the measurable physical quantities determined by these barriers are estimated as a function of the wall
radius: shear strengths and diffusion coefficients for relative sliding of the walls along the axis, as well as frequencies of relative axial
oscillations of the walls. For nonreversible telescopic extension of the walls, maximum overlap of the walls for which threshold static friction
forces are greater than capillary forces is estimated. Possibility of experimental verification of the calculated barriers by measurements of the
estimated physical quantities is discussed.
\end{abstract}

\section{Introduction}

The possibility of relative motion of the walls \cite{cumings00,kis06} in multi-walled carbon nanotubes (MWNTs) makes it promising to use nanotube
walls as movable elements of nanoelectromechanical systems (NEMS) (see \cite{lozovik07} for a review). At present, nanomotors in which the walls of a
MWNT play the role of the shaft and the bush \cite{fennimore03,bourlon04,barreiro08,subramanian07} and memory cells operating on relative sliding of
the walls along the nanotube axis \cite{deshpande06,subramanian10} have been implemented. A number of NEMS have been proposed, which operate on the
relative motion of carbon nanotube walls, including a gigahertz oscillator \cite{zheng02,zheng02a}, Brownian nanomotor \cite{tu05}, a bolt/nut pair
\cite{saito01,lozovik03,lozovik03a} and an accelerometer \cite{wang08,kang09}. Thus the study of relative motion and interaction of carbon nanotube
walls is an actual problem.

However neither the interaction between carbon nanotube walls nor the interaction between graphene layers have been investigated in details. The
measurements of the energy of interaction between graphite layers give the wide scatter of the results (see \cite{benedict98,zacharia04,girifalco56}
and references in \cite{benedict98}). Though experimental value of the interwall interaction energy is available ($23-33$ meV per atom \cite{kis06}),
any measurements of barriers to relative motion of walls of carbon nanotubes are absent up to now. Only upper limits of the shear strength for the
relative motion of the walls along the nanotube axis were measured in a few experiments \cite{cumings00,kis06}. No experimental data on the barriers
to relative motion of graphene layers are available. The value of the critical shear strength for graphite measured in the only known experiment
\cite{soule68} is related to macroscopic structural defects of the graphite sample. Therefore the theoretical studies are particularly valuable for
understanding phenomena and elaboration of NEMS related with relative motion and interaction of carbon nanotube walls.

For existing experimental realizations of relative motion of walls of carbon nanotubes, the movable walls are walls of MWNTs and have relatively
great diameters: $4-10$ nm at the relative sliding of the inner core of the walls attached to nanomanipulator \cite{cumings00}, $7-40$ nm in the
nanomotors \cite{fennimore03,bourlon04,barreiro08,subramanian07}, and $3-7$ nm in the memory cells operating on relative sliding of the walls along
the nanotube axis \cite{deshpande06}. Therefore it is probably that pioneering measurements of the barriers to the relative motion of the walls will
be performed for MWNTs with great diameter. However {\it all previous theoretical considerations} of the barriers to relative motion of walls deal
with double-walled carbon nanotubes (DWNTs) with diameters less than 3 nm
\cite{saito01,kolmogorov00,damnjanovic02,vukovic03,damnjanovic03,belikov04,bichoutskaia05,charlier93,kwon98,palser99,bichoutskaia06,popov09,kolmogorov05}.
That is the considered nanotubes do not correspond to the diameter range for which the experimental support can be expected first. Moreover the DWNTs
with diameter greater than 5 nm have collapsed structure \cite{xiao07}. Therefore, calculations of the barriers to the relative motion of the
adjacent walls of the MWNTs with a wide range of diameters, as well as measurements of these barriers are necessary for further progress in
elaboration of nanotube-based NEMS and for verification of theoretical methods for consideration of interaction between nanotube walls and graphene
layers.

In the present work we study the dependencies of nanotube properties related with the interwall interaction on diameter of the walls. For this
purpose we perform the first calculations of the interwall interaction energies, as well as the barriers to relative sliding of the adjacent walls
along the nanotube axis for a wide range of the wall diameter. The results of these calculations are used to estimate the following physical
quantities: the shear strength for the relative sliding of walls along the nanotube axis, the frequency of the relative axial oscillations and
diffusion coefficient for the short outer wall placed in the middle of the nanotube, the maximum overlap of the inner and outer walls for which the
controlled reversible telescoping can be achieved (the pulled-out core of the inner walls could be completely pushed back by capillary forces
restoring a nanotube to its original retracted condition). The possibilities of experimental study of these quantities is discussed. Comparison of
experimental and calculated values of these quantities can be used as a criterion for the correctness of the results of calculations of the barriers.

The paper is organized as follows. Section II is devoted to analysis of the wall structure and the interwall interaction to prove the choice of pairs
of adjacent walls for the present study. Section III presents the details of computational technique. Section IV gives the results of calculations of
the interwall interaction energies, and the barriers to the relative sliding of the walls, as well as estimations of the physical quantities based on
these barriers. Section V presents discussion of possibility of experimental study of these quantities and summarizes our conclusions.

\section{Structure of wall and interwall interaction}

Neglecting their structure at the ends, which can be either open or closed, carbon nanotubes are single or multiple layers of a cylinder rolled up
from graphene sheets. Only one parameter is needed to fully determine the structure of the middle section of a single-walled carbon nanotube (SWNT)
or a wall of a MWNT: the chirality index $(n, m)$ which corresponds to a two-dimensional lattice vector ${\bf c} = n {\bf a}_1 + m {\bf a}_2$, where
${\bf a}_1$ and ${\bf a}_2$ are the unit vectors of graphene \cite{class,class1}. A segment defined by the vector ${\bf c}$ becomes the circumference
of cylindrical surface of a nanotube wall. Two types of SWNT (nonchiral SWNT) characterized by the chirality index of $(n,n)$ and $(n,0)$ have a
simple translational symmetry, and these are referred to as armchair and zigzag nanotubes forming different pattern of hexagons in circumference.

The adjacent walls of nanotube are commensurate if the ratio of the lengths of their unit cells is a rational fraction. In this case, a pair of walls
can be considered as a quasi-one-dimensional crystal with the length of unit cell equal to the lowest common factor of the lengths of unit cells of
constituent walls. If the length of the wall overlap is considerably greater than the length of the unit cell, barriers to the relative motion of the
commensurate walls are approximately proportional to the length of the wall overlap: $\Delta U=\Delta U_cN_c$, where $\Delta U_c$ is the barrier per
unit cell and $N_c$ is the number of unit cells corresponding to the wall overlap. Conceivably, it is possible to control the barriers to relative
motion of the commensurate walls. Lack of commensurability between the adjacent walls implies a dramatic weakening of the corrugation in the
interwall interaction energy potential surface \cite{kolmogorov00}. In contrast to the commensurate case, barriers to relative motion of
incommensurate walls of do not increase with the length of the wall overlap, but fluctuate near the average value
\cite{lozovik03,lozovik03a,kolmogorov00,damnjanovic02}. These incommensurate systems, even if the wall overlap exceed micron in length, have barriers
to the relative motion of the walls comparable to those with a wall overlap of a single unit cell in length.

For adjacent commensurate walls, at least one of which is chiral, the interwall interaction energy potential surface is extremely flat and
corrugations of the surface are smaller than the accuracy of calculations \cite{vukovic03,belikov04,bichoutskaia05}. This is so because only very
high harmonics of the Fourier transform of the interaction energy $U_a$ between an atom of one wall and the entire other wall make contributions to
the potential barriers due to the incompatibility of the helical symmetries of the walls \cite{damnjanovic02}. Contrary to the case of chiral walls,
{\it ab initio} calculations for the certain DWNTs \cite{bichoutskaia05,charlier93,kwon98,palser99,bichoutskaia06,popov09,kolmogorov05} as well as
calculations using empirical potentials \cite{saito01,kolmogorov00,damnjanovic02,vukovic03,damnjanovic03,belikov04,kolmogorov05} show that DWNTs with
the nonchiral commensurate walls (i.e. both armchair or both zigzag) have considerable barriers to relative motion of walls. Therefore, the pairs of
the nonchiral commensurate walls can be considered as potential components in NEMS for which a precise control of motion or position of the walls is
required. A wide set of such NEMS which operating principle are based on the dependence of the conductivity or/and the interwall interaction energy
on the relative position of the walls on subnanometer scales was proposed. This set includes now a varying nanoresistor
\cite{lozovik03,lozovik03a,lozovik04,yan06}, a stress nanosensor \cite{bichoutskaia06a}, an electromechanical nanothermometer
\cite{bichoutskaia07,popov07}, a nanoactuator for the transformation of a force directed along the nanotube axis to the relative rotation of its
walls \cite{popov07,kuznetsov07}, and a nanoresonator \cite{bichoutskaia09a}. Thus, analysis of the relative motion and interaction of nonchiral
commensurate carbon nanotube walls holds the key to success of these applications. Moreover, due to considerable value of the barriers to relative
motion of walls, nanotubes with such walls should be considered as best candidates for measurements of these barriers.

\section{Computational technique}

Since the system consisting of two nonchiral commensurate walls with great radius has the unit cell which contains up to several thousands atoms,
{\it ab initio} calculations are not available for such a system. Thus we perform empirical calculations of the interwall interaction energy and the
barriers to the relative motion of  the nonchiral commensurate walls with the aim to obtain the qualitative dependencies of these quantities on the
wall radius. The interaction between atoms of adjacent walls is described here by 6-12 Lenard-Jones potential
$U=4\epsilon((\sigma/r)^{12}-(\sigma/r)^{6})$. Note that contribution of many-body terms into the complete Van der Waals interaction is not essential
at relative motion of linear clusters interacting along its full length \cite{kim06}. The parameters of potential, $\epsilon=2.968$ meV and
$\sigma=3.407$ \AA, were fitted to interlayer distance and modulus of elasticity of graphite and used in study of the interwall interaction
energypotential surface of the DWNTs \cite{saito01,lozovik03,lozovik03a,belikov04}. The upper cutoff distance of the potential is chosen so that
calculate the barriers to the relative sliding of the walls with the accuracy 0.1 \%. Namely, the upper cutoff distance is 42, 120, 25 and 35 \AA~
for ($n$,$n$)@($n$+5,$n$+5), ($n$,$n$)@($n$+6,$n$+6), ($n$,0)@($n$+9,0), and ($n$,0)@($n$+10,0) pairs of adjacent walls, respectively. The length of
the outer wall is equal to the length of the unit cell and the length of the inner wall is chosen so that all pairs of atoms with interatomic
distances within the cutoff distance are taken into consideration. The structure of the walls is obtained by mapping of graphene with bond length
$a_0=1.41$ \AA~ on cylindrical surface. The walls are considered as rigid. Account of the deformation of walls is not essential for the shape of
potential surface both for the interwall interaction energy of DWNTs \cite{kolmogorov00} and the intershell interaction energy of carbon
nanoparticles \cite{lozovik00,lozovik02}. For example, the barriers to the relative wall rotation and sliding along the axis for the (5,5)@(10,10)
DWNT calculated for walls with unannealed structure \cite{belikov04} coincide within 14 \% with results of Dresselhaus {\it et al.} (used annealed
wall structure) \cite{saito01}. Note also that 6-12 Lenard-Jones potential was used to study the dependencies of the ground state interwall
interaction energy \cite{bellarosa06,vardanega10} and radial breathing mode frequencies \cite{vardanega10} as function of radius and chiral angle of
walls for DWNTs with chiral walls and to determine characteristics of nanotube-based bolt/nut pairs \cite{salehinia11}.

\begin{figure}
\includegraphics[width=\textwidth]{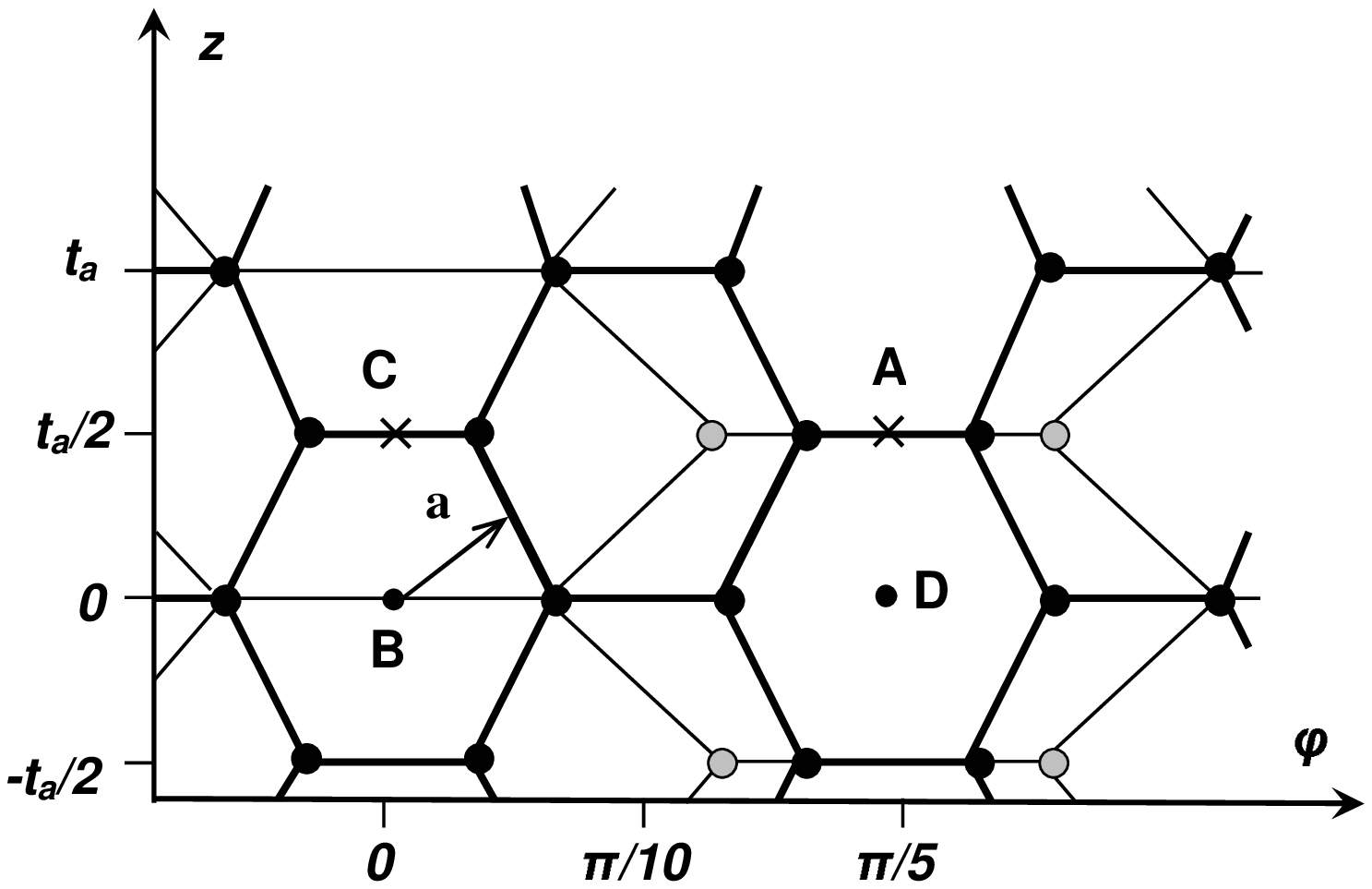}
\caption{The schematic diagram in cylindrical coordinates for the relative positions of walls of the (5,5)@(10,10) DWNT corresponding the maximum of
the interwall interaction energy. Bonds and atoms of the inner (5,5) wall are shown by thin lines and grey circles, respectively. Bonds and atoms of
the outer (10,10) wall are shown by thick lines and black circles, respectively. Point A corresponds to the coincident symmetry axes $U_2$ of the
inner and outer walls which passes through the midpoint of the carbon bond, point B corresponds to the coincident axis $U_2$ of the inner wall which
passes through the midpoint of the carbon bond and axis $U_2$ of the outer wall which passes through the center of the hexagon, point C corresponds
to the coincident axis $U_2$ of the inner wall which passes through the center of the hexagon and axis $U_2$ of the outer wall which passes through
the midpoint of the carbon bond, point D corresponds to the coincident axes $U_2$ of the inner and outer walls which passes through the center of
hexagon. Vector ${\bf a}=(\delta_z/2,\delta_{\phi}/2)$ shows the displacement and rotation of one of the walls to the relative position corresponding
to the minimum of the interwall interaction energy.} \label{fig:1}
\end{figure}

According to the classification scheme of DWNTs with commensurate wall \cite{belikov04} (which is also the classification scheme of pairs of coaxial
commensurate walls) such pairs form families with the same interwall distance and unit cell length for each pair of any family. Here we consider here
four families of such pairs: ($n$,$n$)@($n$+5,$n$+5), ($n$,$n$)@($n$+6,$n$+6), ($n$,0)@($n$+9,0), and ($n$,0)@($n$+10,0) with the interwall distances
3.37, 4.04, 3.50 and 3.89 \AA, respectively. The inner wall chirality index $n$ up to 250 are considered, such chirality index corresponds to the
outer wall radiuses from 0.65 to 17 nm for the pairs of armchair walls and from 0.7 to 10 nm for the pairs of zigzag walls.

To analyze the nanomechanical properties connected with the relative motion and interaction of nanotube walls, we must calculate the ntershell
interaction energy potential surface, i.e. the dependence of the energy $U$ of the interaction between two adjacent walls on the coordinates
describing the relative position of the walls (angle $\phi$ of the relative rotation of the walls about the nanotube axis and length $z$ of the
relative displacement of the walls along this axis).

The relative positions of the walls corresponding to a higher symmetry of the system are critical points of interwall interaction energy $U(z,\phi)$
(extrema or saddle points) \cite{damnjanovic99}. For such relative positions of the walls, some of the second-order axes $U_2$ of the inner and outer
walls are in line. The second-order $U_2$ axis is perpendicular to the principal axis of the wall and passes through the midpoint of the carbon bond
or the center of the hexagon. The detailed consideration for the case of coaxial nonchiral commensurate walls is presented in
\cite{damnjanovic03,bichoutskaia06}. In particular, the possible relative positions corresponding to the critical points are listed in
\cite{bichoutskaia06}, and the relative positions corresponding to the minima of the interwall interaction energy are determined using 6-12
Lenard-Jones potential in \cite{damnjanovic03}. In accordance with the symmetry of armchair $(n_1,n_1)@(n_2,n_2)$ and zigzag $(n_1,0)@(n_2,0)$ DWNTs,
the unit cell of the interwall interaction energy potential surface of such pairs of walls is a rectangle with sides $\delta_z=t_a/2$ and
$\delta_\phi=\pi N/n_1n_2$, where $t_a$ is the length of the unit cell of the wall and $N$ is the greatest common division of numbers $n_1$ and $n_2$
\cite{damnjanovic99}. As an example we present in figure 1 the schematic diagram describing the relative positions of walls of the (5,5)@(10,10) DWNT
corresponding to the maximum of the interwall interaction energy.

Empirical calculations with the use of the Lennard-Jones potential show that for all considered armchair ($n$,$n$)@($n$+5,$n$+5) DWNT with $n=5-15$
and zigzag ($n$,0)@($n$+9,0) DWNT with $n=9-18$, the interwall interaction energy can be successfully interpolated within an accuracy of about 1\% of
the value of the barriers by the following expression \cite{belikov04}:

\begin{equation}\label{expansion2}
               U(\phi,z)=U_0-\frac{\Delta U_{\phi}}{2} \cos \left(\frac{2\pi}{\delta_{\phi}}
\phi\right)-\frac{\Delta U_z}{2} \cos \left(\frac{2\pi}{\delta_z} z\right),
\end{equation}

\noindent where $U_0$ is the mean energy of interwall interaction, $\Delta U_z$ and $\Delta U_{\phi}$ are the barriers to the relative sliding of the
walls along the nanotube axis and the relative rotation of the walls (between their relative positions corresponding to the interaction energy
minimum), respectively. Henceforth, we will normalize the values of $U_0$, and $\Delta U_z$ to an atom of the outer wall. This empirical results is
confirmed for several DWNTs with the use of density functional method within the accuracy of the calculations
\cite{bichoutskaia05,popov09,bichoutskaia09}. In this case, where the expression (\ref{expansion2}) is adequate, the four types of the critical
points on the surface are the global minimum, global maximum, and two saddle points. Thus to determine the barriers to the relative motion of the
walls the interwall interaction energy is calculated here at the critical points. For DWNTs with compatible rotational symmetry of the walls ($N\neq
1$), the barrier to the relative rotation, $\Delta U_{\phi}$, can be significant. For DWNTs with incompatible rotational symmetry of the walls
($N=1$), both {\it ab initio} \cite{bichoutskaia06} and empirical \cite{vukovic03,belikov04} results show that the dependence of the interwall
interaction energy on the angle $\phi$ is extremely flat, thus the second term in expansion (\ref{expansion2}) can be neglected. The present work is
devoted to pairs of adjacent walls with great radius. The rotational symmetry of the walls is incompatible for all such pairs. Therefore only the
barriers to the relative sliding of the walls along the nanotube axis and physical quantities related with these barriers are calculated.

\section{Properties related with interwall interaction}

\begin{figure}[htb]
\includegraphics[width=\textwidth]{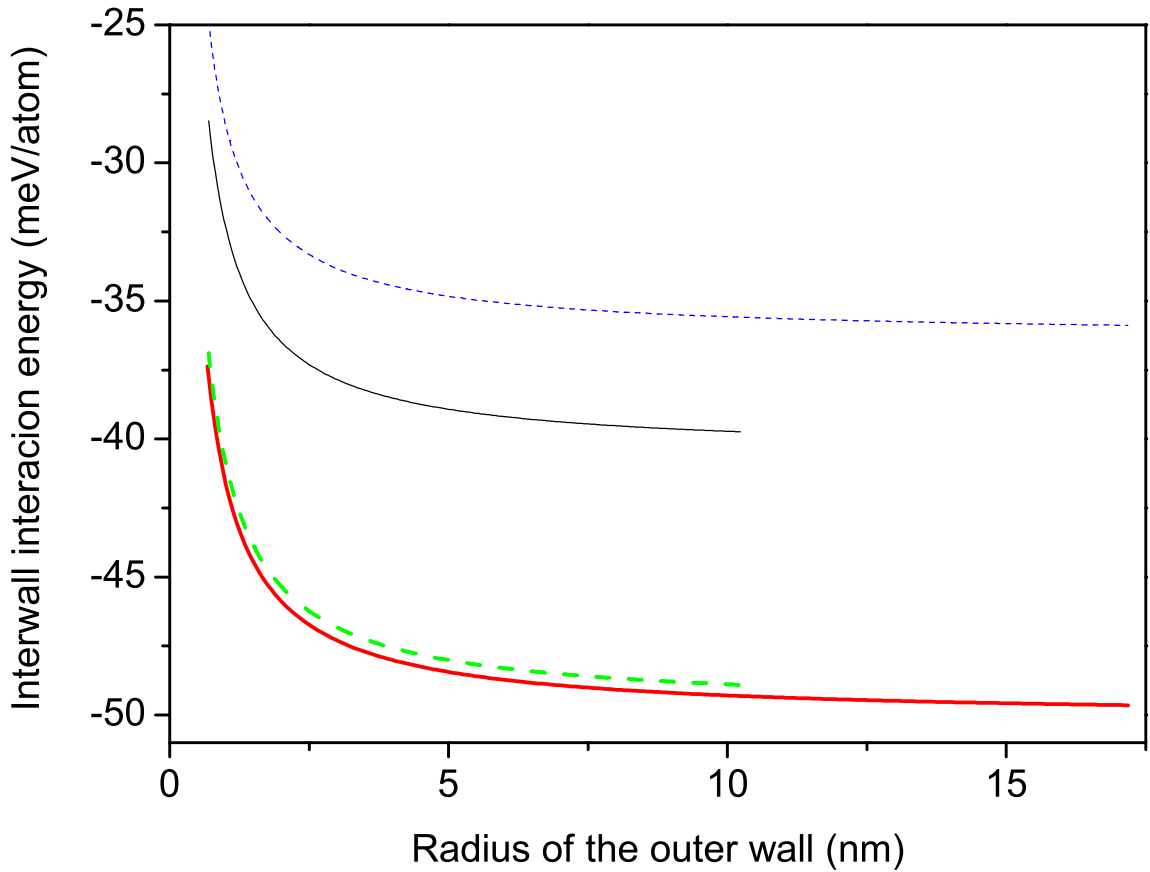}
\caption{Calculated interwall interaction energy (in meV per atom of the outer wall) as a function of the radius of the outer wall (in nm) of the
($n$,$n$)@($n$+5,$n$+5) (red solid thick line), ($n$,$n$)@($n$+6,$n$+6) (blue dotted thing line), ($n$,0)@($n$+9,0) (green dashed thick line), and
($n$,0)@($n$+10,0) (black solid thing line) families of wall pairs.} \label{fig:2}
\end{figure}

The calculated dependencies of the interwall interaction energy corresponding to the minimum of the function $U(z,\phi)$ on the radius of the outer
wall are presented in figure 2 for four mentioned above families of the adjacent wall pairs. These values of the interwall interaction energy lie in
the range $24-50$ meV per atom of the outer wall. This result is in good agreement with the experimental values of the energy of interaction between
graphite layers 52$\pm$5 meV/atom obtained recently from the experiments on the thermal desorption of polyaromatic hydrocarbons from a graphitic
surface \cite{zacharia04}, 43 meV/atom from the heat-of-wetting experiment \cite{girifalco56}, 35$\pm$10 meV/atom from an analysis of the structure
of collapsed carbon nanotubes \cite{benedict98}, and between the walls of MWNTs 20$-$33 meV/atom \cite{kis06}. The figure 2 shows that for pairs from
the same family the interwall interaction energy tends to a constant value with increase of the outer wall radius. We found that for the great
radiuses of the outer walls the dependencies of the interwall interaction energy on the outer wall radius can be interpolated by the following
expression:

\begin{equation}\label{ur}
U = U_\infty+U_1\exp\left(-\frac{R_2}{R'}\right),
\end{equation}

\noindent where $U_\infty$, $U_1$ and $R'$ are the fitting parameters, $R_2$ is the radius of the outer wall. The calculated values of the fitting
parameters are listed in table 1. Note that the fitting parameter $R'$, which characterizes the rate of exponential decay of the interpolated
dependencies, coincides within the calculation accuracy for both considered families of the armchair adjacent wall pairs of and for both considered
families of the zigzag adjacent wall pairs. The parameter $U_1$ is order of magnitude less than the parameter $U_\infty$ for all considered families
of the pairs of the adjacent wall. Therefore the interwall interaction energy per one atom of the outer wall only slightly increases with the radius
increase and tends to a constant value $U_\infty$ for these pairs of walls with the radius greater than 5 nm.

\vspace{0.5cm} {\bf Table 1} Calculated fitting parameters $U_\infty$, $U_1$ and $R'$ of the interpolation by expression (\ref{ur}) of the
dependencies of the interwall interaction energy on the radius of the outer wall. The interpolation is performed for the radius greater than 5 nm.

\vspace{0.1cm}
\begin{tabular}{|l|c|c|c|}
\hline
family of wall pairs & $U_\infty$ (meV/atom) & $U_1$ (meV/atom) & $R'$ (nm) \\
\hline
($n$,$n$)@($n$+5,$n$+5) & -49.713$\pm$0.004 & 3.65$\pm$0.03 & 4.61$\pm$0.04 \\
($n$,$n$)@($n$+6,$n$+6) & -35.944$\pm$0.003 & 3.18$\pm$0.03 & 4.64$\pm$0.04 \\
($n$,0)@($n$+9,0) & -49.132$\pm$0.003 & 4.87$\pm$0.03 & 3.40$\pm$0.02 \\
($n$,0)@($n$+10,0) & -39.942$\pm$0.004 & 4.43$\pm$0.02 & 3.39$\pm$0.02 \\
\hline
\end{tabular}

\vspace{0.5cm}We confirm also the previous qualitative results of atomistic empirical \cite{saito01,belikov04,bellarosa06,vardanega10} and {\it ab
initio} calculations \cite{bichoutskaia06} for DWNTs with the smaller radiuses: 1) the interwall interaction energy per one atom increases with an
increase of the radius of DWNT \cite{belikov04,bichoutskaia06}; 2) the interwall interaction energy per one atom is greater for DWNTs with the
interwall distance close to the distance between graphite layers 3.4 \AA~ \cite{saito01,bichoutskaia06,bellarosa06,vardanega10}.

\begin{figure}[htb]
\includegraphics[width=\textwidth]{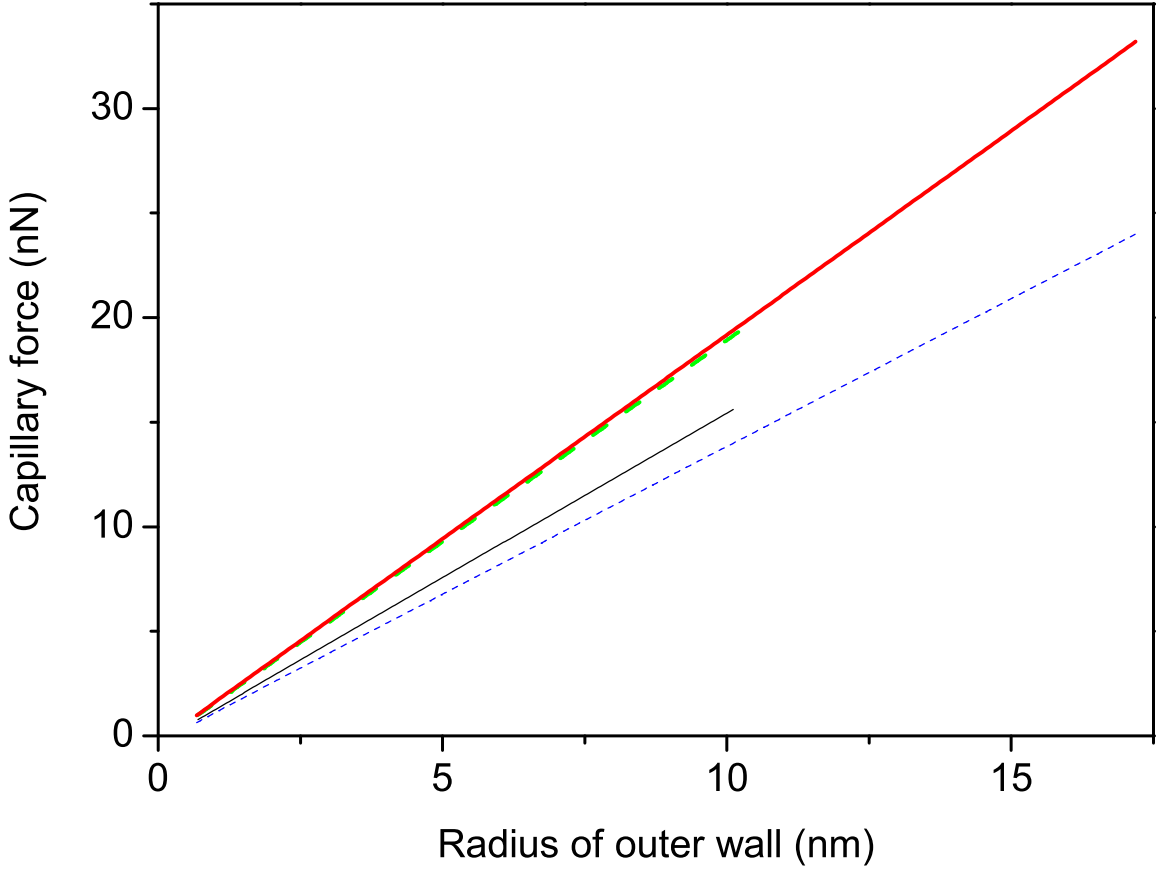}
\caption{Calculated capillary force pulling the inner wall back into the outer wall (in nN) as a function of the radius of the outer wall (in nm) of
the ($n$,$n$)@($n$+5,$n$+5) (red solid thick line), ($n$,$n$)@($n$+6,$n$+6) (blue dotted thing line), ($n$,0)@($n$+9,0) (green dashed thick line),
and ($n$,0)@($n$+10,0) (black solid thing line) families of wall pairs.} \label{fig:3}
\end{figure}

The analogous tendency to a constant of the interwall interaction energy per one atom with the increase of wall radius was obtained in the framework
of approach with continuous distribution of atom density not only for interaction between pairs of adjacent walls but also for the case where the
interaction with the next walls of the inner core or outer shell of the MWNT is taken into account \cite{zheng02a}. That is the account of this
additional interaction energy leads only to increase of the parameter $U_\infty$ and this increase is found to be within 27 \% for any number of
walls in the MWNT \cite{zheng02a}. Such a change of the parameter $U_\infty$ is comparable to the scatter in the listed above experimental data for
graphite interlayer interaction energy. Therefore we believe that the corrections to the interaction energy due to the interaction with the next
walls of the inner core or outer shell of the MWNT obtained in the framework of the approach with continuous distribution of atom density
\cite{zheng02a} can be applicable also to the our the results obtained in the framework of the atomistic approach within the accuracy of both
approaches. Thus the atomistic calculations of the mentioned above corrections to the interaction energy is beyond the purpose of the present work
which is devoted mainly to the physical properties related with the barriers to the wall motion.

\begin{figure}[htb]
\includegraphics[width=\textwidth]{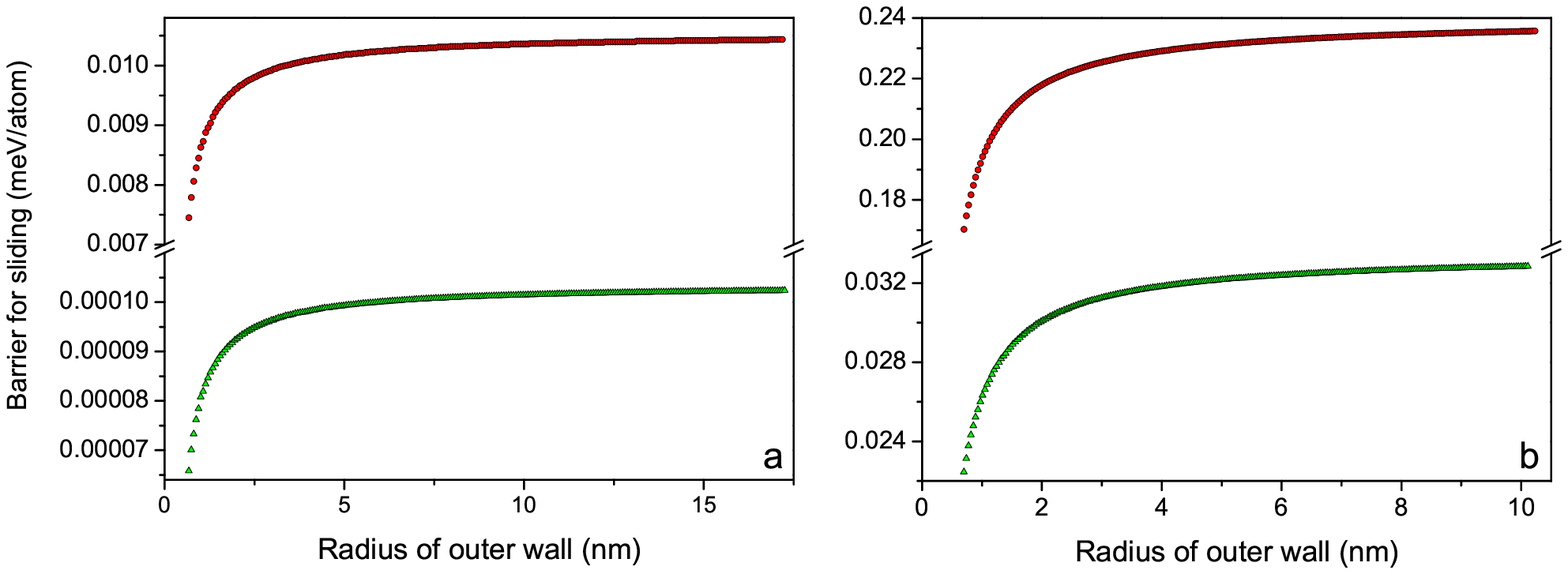}
\caption{Calculated barrier to the relative sliding of the walls (in meV per atom of the outer wall) as a function of the radius of the outer wall
(in nm). (a) the ($n$,$n$)@($n$+6,$n$+6) family of wall pairs (green triangles) and the ($n$,$n$)@($n$+5,$n$+5) family of wall pairs (red circles),
(b) the ($n$,0)@($n$+10,0) family of wall pairs (green triangles) and the ($n$,0)@($n$+9,0) family of wall pairs (red circles).} \label{fig:4}
\end{figure}

A telescopic extension of the inner wall gives rise of a capillary force $F_c$ pulling the inner wall back into the outer wall. The average value of
this force is defined as

\begin{equation}\label{forcecap}
\langle F_c\rangle=\left\langle\frac{d U}{d L_{ov}}\right\rangle=\frac{U_04n_2}{t_a},
\end{equation}

\noindent where $L_{ov}$ is the overlap length of the walls, $4n_2$ is the number of atoms in the unit cell of the outer wall. The calculated
dependencies of the average capillary force on the radius of the outer wall are presented in figure 3. Since the interwall interaction energy tends
to a constant value for the pairs of walls from the same family with the increase of the wall radius, the average capillary force for such pairs is
approximately proportional to the radius.

The calculated dependencies of the barriers $\Delta U_z$ for the relative sliding of walls on the outer wall radius are shown in figure 4 for all
considered families of the wall pairs. The relatively small increase of the interwall distance leads to decrease of of the barrier to the relative
sliding of the walls by orders of magnitude (compare ($n$,$n$)@($n$+5,$n$+5) and ($n$,$n$)@($n$+6,$n$+6), ($n$,0)@($n$+9,0) and ($n$,0)@($n$+10,0)
families). Therefore the contribution of the interaction between the other walls of the MWNT into the barrier to the relative sliding of the walls is
negligible in comparison with the contribution of the interaction between the adjacent walls. Note that the barrier to relative sliding of walls of
(5,5)@(10,10) pair of walls calculated here is close to that calculated using density functional method \cite{popov09}. Analogously to the
dependencies of the interwall interaction energy on the outer wall radius, the dependencies of the barriers to relative sliding of walls for the
great radius of the outer wall are interpolated by the following expression

\begin{equation}\label{dur}
\Delta U_z = \Delta U_\infty+\Delta U_1\exp\left(-\frac{R_2}{R'_b} \right),
\end{equation}

\noindent where $\Delta U_\infty$, $\Delta U_1$ and $R'_b$ are the fitting parameters. The calculated values of the fitting parameters for families
of wall pairs with greater barriers ($n$,$n$)@($n$+5,$n$+5), ($n$,0)@($n$+9,0), and ($n$,0)@($n$+10,0) are listed in table 2. Analogously to the
dependencies of the interwall interaction energy on the radius of the outer wall, the parameter $\Delta U_1$ is order of magnitude less than the
parameter $\Delta U_\infty$ for all families of the wall pairs presented in table 2. Therefore for these pairs with the outer wall radius greater 5
nm interwall interaction energy per one atom of the outer wall only slightly increases with the radius increase and tends to a constant value $\Delta
U_\infty$.

\vspace{1cm}{\bf Table 2} Calculated fitting parameters $\Delta U_\infty$, $\Delta U_1$ and $R'_b$ of the interpolation by expression (\ref{dur}) of
the dependencies of the barrier to relative sliding of the walls on the radius of the outer wall. The interpolation is performed for the radius
greater than 5 nm.

\vspace{0.1cm}\begin{tabular}{|l|c|c|c|} \hline
family of wall pairs & $\Delta U_\infty$ (meV/atom) & $\Delta U_1$ (meV/atom) & $R'_b$ (nm) \\
\hline
($n$,$n$)@($n$+5,$n$+5) & 0.0104 & 0.0010 & 3.48$\pm$0.18 \\
($n$,0)@($n$+9,0) & 0.237 & 0.013 & 5.83$\pm$0.03 \\
($n$,0)@($n$+10,0) & 0.0330 & 0.0020 & 5.95$\pm$0.07 \\
\hline
\end{tabular}

\vspace{0.5cm}The calculated values of barriers $\Delta U_z$ for the relative sliding of the adjacent walls can be used for calculating the values of
a number of physical quantities characterizing the interaction and relative motion of the nanotube walls. Expression (\ref{expansion2}) for the
dependence of the interwall interaction energy $U(\phi,z)$ on the relative position of the walls defines the threshold static friction force $F_z$
for the relative sliding of the walls along the nanotube axis:

\begin{equation}\label{fx}
F_z = \frac {4n_2 \pi L_{ov}\Delta U_{z} }{\delta_z t_a}.
\end{equation}

The shear strength for relative sliding of the walls along the nanotube axis is defined as

\begin{equation}\label{m}
M_z=\frac {F_z} S,
\end{equation}

\noindent where $S$ is the area of the overlap surface of the walls,

\begin{equation}\label{square}
S=2\pi L_{ov}\left(\frac{R_1+R_2}2\right)=\pi L_{ov}(R_1+R_2),
\end{equation}

\noindent where $R_1$ is the radius of the inner wall. Figure 5 presents the dependencies of the shear strength on the radius of the outer wall for
four considered families of the wall pairs.

\begin{figure}[htb]
\includegraphics[width=\textwidth]{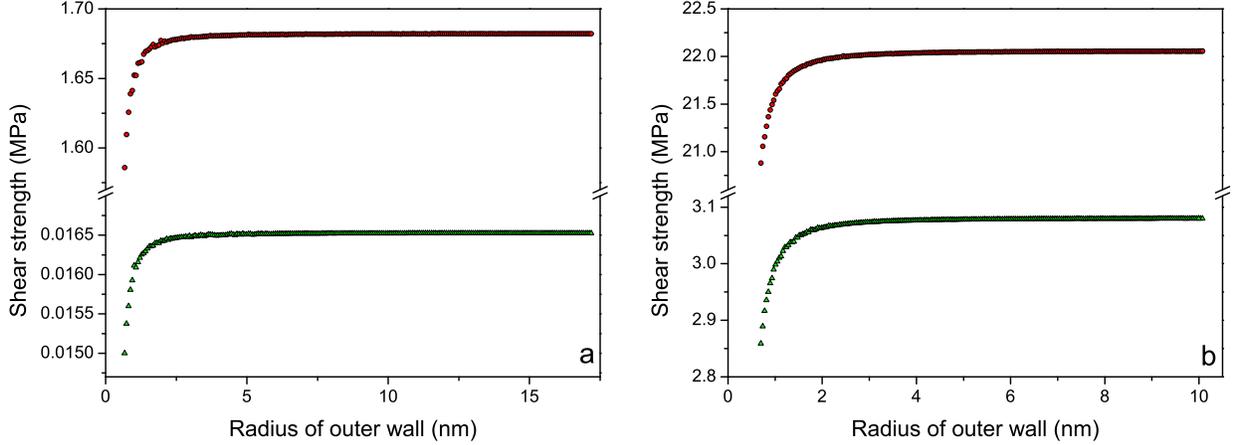}
\caption{Calculated shear strength for relative sliding of the walls (in MPa) as a function of the radius of the outer wall (in nm). (a) the
($n$,$n$)@($n$+6,$n$+6) family of wall pairs (green triangles) and the ($n$,$n$)@($n$+5,$n$+5) family of wall pairs (red circles), (b) the
($n$,0)@($n$+10,0) family of wall pairs (green triangles) and the ($n$,0)@($n$+9,0) family of wall pairs (red circles).} \label{fig:5}
\end{figure}

\begin{figure}[htb]
\includegraphics[width=\textwidth]{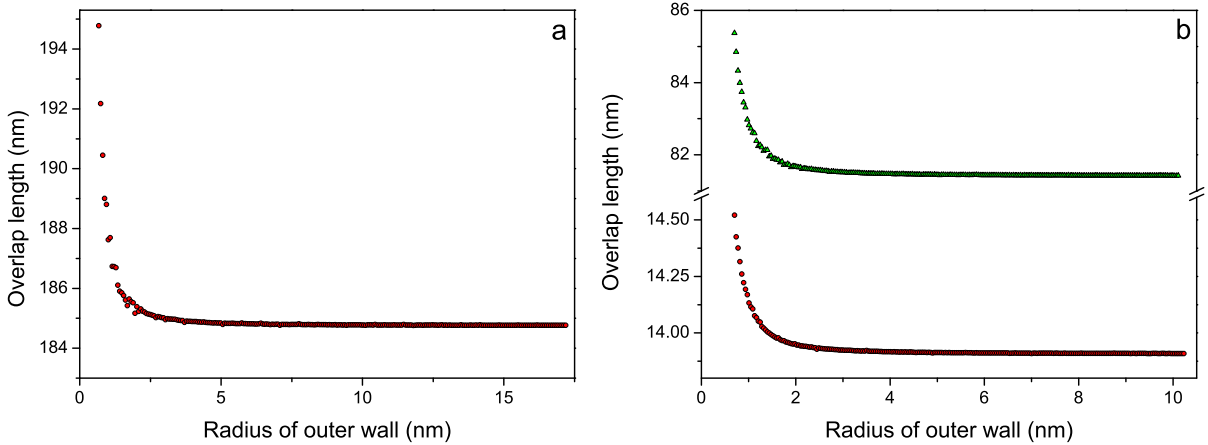}
\caption{Calculated minimal overlap length of the walls for which the static friction force prevents the core of the inner walls from being pulled by
the capillary force (in nm) as a function of the radius of the outer wall (in nm). (a) the ($n$,$n$)@($n$+5,$n$+5) family of wall pairs (circles),
(b) the ($n$,0)@($n$+10,0) family of wall pairs (green triangles) and the ($n$,0)@($n$+9,0) family of wall pairs (red circles).} \label{fig:6}
\end{figure}

The equality of forces $F_c=F_z$ acting on the movable wall controls the minimal overlap length $L_m$ of the walls for which the static friction
force $F_z$ prevents the inner wall from being pulled by the capillary force $F_c$. Substituting Eqs. (\ref{forcecap}) and (\ref{fx}) for forces
$F_c$ and $F_z$, respectively, in this equality  we have

\begin{equation}\label{lm}
L_m=\frac{\delta_z U_0}{\pi \Delta U_z}.
\end{equation}

\noindent For the ($n$,$n$)@($n$+6,$n$+6) family the calculated value of the minimal overlap length $L_m$ is about 14 $\mu$m which is too much for
nanotubes produced. The dependencies of the minimal overlap length $L_m$ on the radius $R_2$ of the outer wall for other considered families of the
wall pairs are presented in figure 6.

\begin{figure}[htb]
\includegraphics[width=\textwidth]{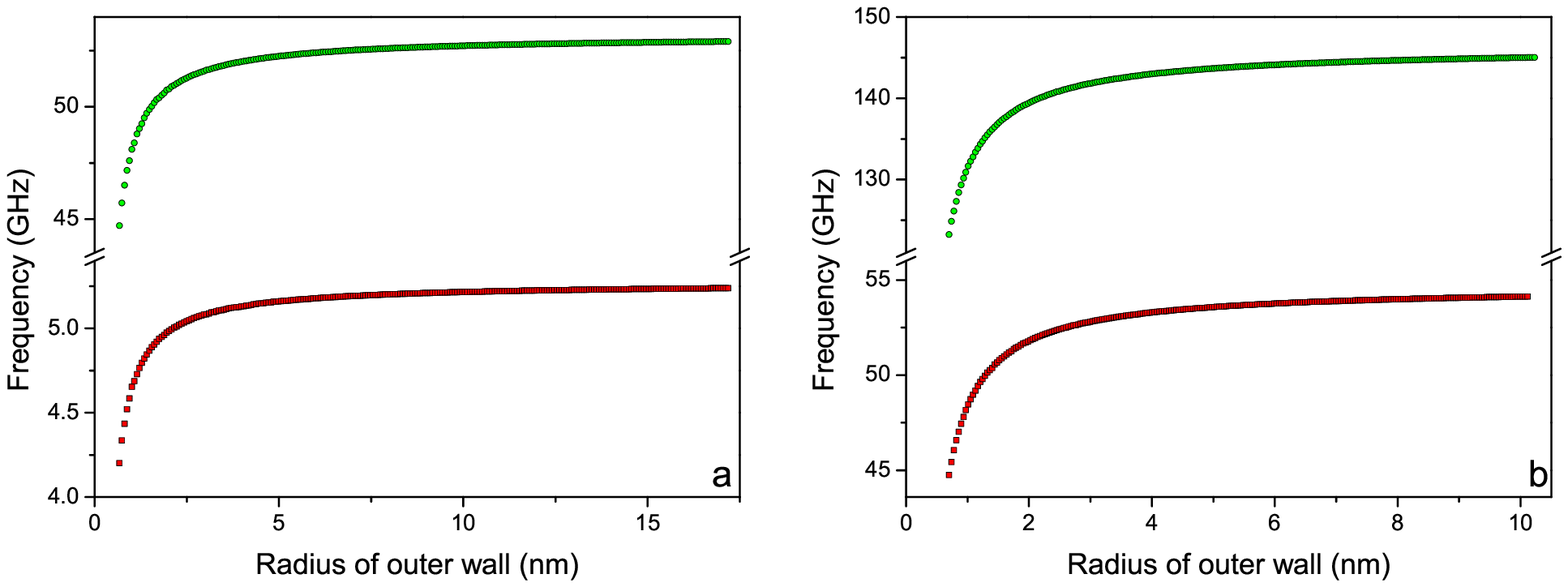}
\caption{Calculated frequencies of relative axial oscillations of the outer wall (in GHz) as a function of the radius of the outer wall (in nm). (a)
the ($n$,$n$)@($n$+5,$n$+5) family of wall pairs (green circles), the ($n$,$n$)@($n$+6,$n$+6) family of wall pairs (red squares); (b) the
($n$,0)@($n$+9,0) family of wall pairs (green circles), the ($n$,0)@($n$+10,0) family of wall pairs (red squares).} \label{fig:7}
\end{figure}

Let us consider the case where the outer wall is shorter than adjacent wall and placed near middle of the adjacent wall without telescopic extension
so that the distance between the edges of the outer wall and the adjacent wall is considerably greater than the parameter $\sigma$ of the potential.
In such a case influence of the wall edges on the interwall interaction energy can be disregarded. To find the frequency of small axial oscillations
of the outer wall in this case, we approximate expression (\ref{expansion2}) for the dependence of the interwall interaction energy on their relative
position by a parabolic potential well in the vicinity of the minimum $U_m$ of this dependence,

\begin{equation}\label{expasion2zf}
U(z')=U_m+\frac{k_zz'^2}{2},
\end{equation}

\noindent where $z'$ is the wall displacement along the nanotube axis relative to the energy minimum of interwall interaction. For
$(n_1,n_1)@(n_2,n_2)$ and $(n_1,0)@(n_2,0)$ pairs of walls, we have

\begin{equation}\label{k}
k_z=\frac{8 n_2 L_{ov} \Delta U_z}{t_a} \left( \frac {2\pi}{t_a} \right)^2.
\end{equation}

Frequency $\nu_z$ of small relative axial oscillations of the wall is defined as

\begin{equation}\label{freq}
\nu_z=\frac1{2\pi}\sqrt{\frac{k_z}{M}},
\end{equation}

\noindent where $M$ is the mass of the outer wall, $M=4n_2m_0L/t_a$, where $L$ is the length of the outer wall. The expression for the frequency of
the axial oscillations of the outer wall assumes the form

\begin{equation}\label{frec_z}
\nu_z=\frac1{t_a}\sqrt{\frac{2\Delta U_{z}}{m_0}}.
\end{equation}

\noindent Figure 7 shows the dependencies of this frequency on the radius of the outer wall for four considered families of wall pairs.

\begin{figure}[htb]
\includegraphics[width=\textwidth]{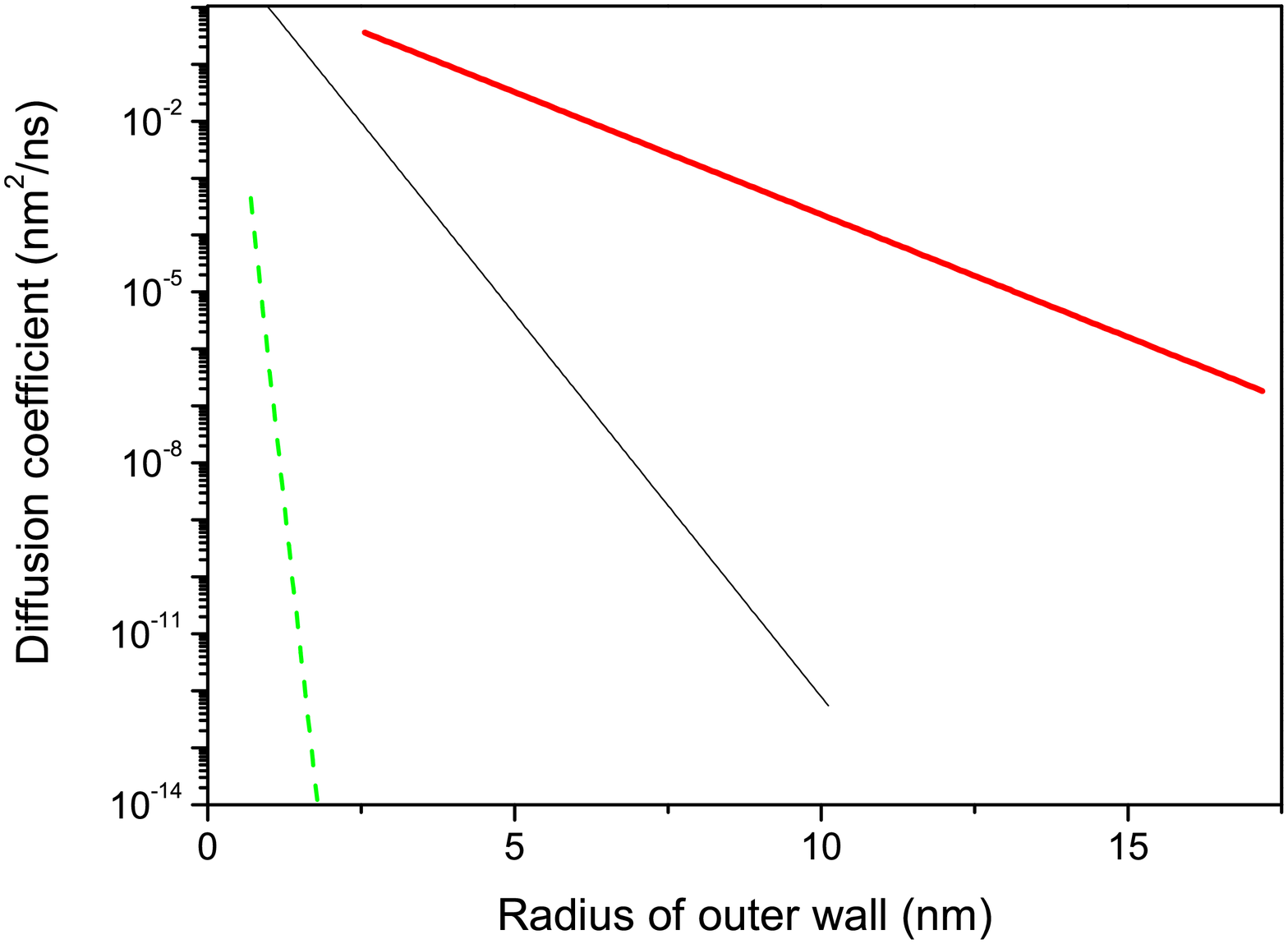}
\caption{Calculated diffusion coefficient of outer wall of length 10 nm (in nm$^2$/ns) as a function of the radius of the outer wall (in nm) of the
($n$,$n$)@($n$+5,$n$+5) (red solid thick line), $n$,0)@($n$+9,0) (green dashed thick line), and ($n$,0)@($n$+10,0) (black solid thing line) families
of wall pairs.} \label{fig:8}
\end{figure}

If the energy of thermal motion of a short outer wall is much lower than the barrier $\Delta U$ to the sliding of this wall along the nanotube axis
($k T \ll \Delta U$, where $\Delta U=4n_2\Delta U_zL/t_a$), then hopping diffusion or drift of the shorter wall along the nanotube axis takes place.
Recently such a drift actuated by imposing of thermal gradient was observed for short shell consisting of several outer walls of many-walled carbon
nanotube \cite{barreiro08}. The Fokker-Planck equation has been obtained in a general case of diffusion and drift of the walls along the helical line
\cite{lozovik03,lozovik03a}. In the case of the motion of the mobile wall along the nanotube axis diffusion coefficient $D_z$ is defined as

\begin{equation}\label{dz}
                D_z =  \frac{1}{2} \Omega_z \delta_z^2 \exp \left( - \frac{4n_2\Delta U_zL}{kTt_a} \right)
\end{equation}

\noindent where $\Omega_z$ is the preexponential factor in the Arrhenius formula for the frequency of the jump of the mobile wall between equivalent
minima of the interwall interaction energy.

Simulation based on the molecular dynamics method shows that the preexponential factor in the Arrhenius formula for the frequency of the jump of a
C$_{60}$@C$_{240}$ nanoparticle shell between equivalent minima of intershell interaction energy is $\Omega=540\pm180$ GHz \cite{lozovik00}. We found
that frequency of small relative rotational oscillations near the minima of intershell interaction energy calculated here for such a nanoparticle is
$\nu=60$ GHz for the same shape of the shells and the same potential of interaction between atoms of the shells. Thus, ratio $\Omega/\nu$ of the
preexponential factor in the Arrhenius formula to the frequency of small oscillations is approximately 10. We believe that ratio $\Omega/\nu$ for
other carbon nanostructures with enclosed graphene layers (in particular, carbon nanotubes) is on the same order of magnitude. We used the value of
$\Omega_z/\nu_z\sim10$ for estimating the diffusion coefficients for diffusion of the outer wall along the nanotube axis. For the
($n$,$n$)@($n$+6,$n$+6) family of the wall pairs the ratio $\Delta U/kT>0.1$ for the short outer wall of length 100 nm for all considered pairs of
walls. Thus the expression (\ref{dz}) is not adequate for estimation of the diffusion coefficient $D_z$ for this family of the wall pairs. The
calculated dependencies of the diffusion coefficients for the motion of the short outer wall 10 nm in length moving along the infinite inner wall on
the radius of the outer wall are presented in figure 8 for other three considered families of the wall pairs. Only pairs with $\Delta U/kT<0.1$ are
shown. The lower limit $D_z=10^{-14}$ nm$^2$/ns of the diffusion coefficient range shown in figure 8 corresponds to displacement of the outer wall by
$\sim 1$ nm during a day.

Under the action of force $F_d$ directed along the nanotube axis, the short outer wall may drift along the axis at a velocity $v=B_zF_d$, where $B_z$
is the mobility of the outer wall for its motion along the axis. In the case of a small force, $F_d\delta_z \ll 2\Delta U$, mobility $B_z$ is related
to the diffusion coefficient by the Einstein relation $D_z=kTB_z$ \cite{lozovik03,lozovik03a}.

\section{Discussion and conclusions}

The dependencies of interwall interaction energies, as well as barriers to relative sliding of the walls along the axis, on the radius of the outer
wall are calculated for the pairs of adjacent nonchiral commensurate walls of carbon nanotubes with the wide range of radiuses. It is found that for
walls with radius greater than 5 nm both the interwall interaction energy and barriers to relative sliding of the walls per one atom of the outer
wall only slightly increase with the radius increase and tend to the constant values. The results of these calculations are used to obtain for these
pairs of walls the dependencies on the radius of the outer wall for the wide set of the measurable physical quantities related to the calculated
barriers. Certain of the calculated physical quantities which are determined only by the barrier to sliding per one atom of the outer wall and do not
depend on the radius of the outer wall also only slightly increase with the radius increase and tend to the constant values. Such quantities are the
shear strength for the relative sliding of the walls along the nanotube axis, the frequency of the axial oscillations of the outer wall and the
maximum overlap of the inner core and outer shell of the walls for which the controlled reversible telescoping can be achieved (the pulled-out core
could be completely pushed back by capillary forces restoring the nanotube to its original retracted condition). The diffusion coefficients for the
relative sliding of the walls along the axis of the nanotube is determined by the total barrier to the relative sliding and exponentially decrease
with the radius increase.

Let us consider how to experimentally verify our results. At present, only the upper boundary of the shear strength for the relative sliding of the
walls along the nanotube axis is determined with the help of atomic force microscopy is $M_z<0.04$ MPa \cite{kis06}. In most cases, the barriers to
the sliding of adjacent walls (and, hence, the corresponding shear strength) are negligibly small (due to the incompatibility of translational
symmetry of the walls for incommensurate walls \cite{damnjanovic02} and to incompatibility of helical symmetries of the walls for commensurate chiral
walls \cite{kolmogorov00,belikov04,bichoutskaia05}). In experiment \cite{kis06}, the shear strength was measured only for a single pair of walls of a
MWNT, which was characterized by a smaller value of this strength as compared to other pairs of adjacent walls. The chirality indices of the walls
were not determined. Thus, we believe that the experimental value of the upper boundary of the shear strength for the relative sliding of the walls
corresponds to incommensurate or commensurate chiral walls. However, the barrier to the relative sliding of adjacent walls may attain an appreciable
value only for commensurate nonchiral walls \cite{kolmogorov00,damnjanovic02,vukovic03,belikov04}. The shear strength calculated here for such pairs
of walls exceed by several orders of magnitude the corresponding boundary of this value (0.04 MPa) measured by an atomic force microscope for the
relative sliding of adjacent walls with indeterminate chirality indices. Consequently, the shear strength for these pairs of walls can easily be
determined experimentally. In our opinion, the minimal overlap length of the walls in the case of the telescopic extension of the inner core of the
walls, for which the static friction force prevents the inner core from being pulled by the capillary force, can also be measured using atomic force
microscopy.

The nanotube with a and short outer wall can be obtained via the electrical-breakdown technique \cite{collins01,collins01a}, which consists of
passing a large current through the nanotube and was used for fabrication of nanotube-based nanomotors \cite{bourlon04,barreiro08,subramanian07}.
Such a DWNT with the short outer wall can be used for measurements of the frequency of axial oscillation of the outer wall relative the fixed inner
core of wall using terahertz spectroscopy or Raman spectroscopy. It should be noted that Raman spectra have recently been obtained for isolated
nanotubes \cite{zhang08}. The diffusion coefficient for the short outer wall moving along the nanotube axis can be measured with the help of scanning
electron microscopy which was used to study the motion of the short outer wall driven by thermal gradient \cite{barreiro08}.

Thus we believe that the calculated physical quantities which are determined by the barriers to relative sliding of walls along nanotube axis can be
measured by contemporary experimental techniques. Such measurements will give the first results for verification of the theoretical methods for
studying the interwall interaction in nanotubes and would facilitate progress in developing nanoelectromechanical systems based on the relative
motion of the walls of nanotubes.

\section*{Acknowledgments}

This work has been partially supported by the Russian Foundation of Basic Research (grants 11-02-00604-a and 10-02-90021-Bel)


\begin{thebibliography}{99}
\bibitem{cumings00} J. Cumings, A. Zettl, Science 289 (2000) 602.
\bibitem{kis06} A. Kis, K. Jensen, S. Aloni, W. Mickelson, A. Zettl, Phys. Rev. Lett. 97 (2006) 025501.
\bibitem{lozovik07} Yu.E. Lozovik, A.M. Popov, Physics-Uspekhi, 50 (2007) 749.
\bibitem{fennimore03} A.M. Fennimore, T.D. Yuzvinsky, W.Q. Han, M.S. Fuhrer, J. Cumings, A. Zettl, Nature 424 (2003) 408.
\bibitem{bourlon04} B. Bourlon, D.C. Glatti, L. Forr\'{o}, A. Bachtold, Nano Lett. 4 (2004) 709.
\bibitem{barreiro08} A. Barreiro, R. Rurali, E.R. Hernandez, J. Moser, T. Pichler, L. Forr\'{o}, A. Bachtold, Science 320 (2008) 775.
\bibitem{subramanian07} A. Subramanian, L.X. Dong, J. Tharian, U. Sennhauser, B.J. Nelson, Nanotechnology 18 (2007) 075703.
\bibitem{deshpande06} V. V. Deshpande, H.-Y. Chiu, H. W. Ch. Postma, C. Mik\'{o}, L. Forr\'{o}, M. Bockrath, Nano Lett. 6 (2006) 1092.
\bibitem{subramanian10} A. Subramanian, L.X. Dong, B.J. Nelson, A. Ferreira, Appl. Phys. Lett. 96 (2010) 073116.
\bibitem{zheng02} Q. Zheng, Q. Jiang, Phys. Rev. Lett. 88 (2002) 045503.
\bibitem{zheng02a} Q. Zheng, J.Z. Liu, Q. Jiang, Phys. Rev. B 65 (2002) 245409.
\bibitem{tu05} Z. C. Tu, X. Hu, Phys. Rev. 72 (2005) 033404.
\bibitem{saito01} R. Saito, R. Matsuo, T. Kimura, G. Dresselhaus, M. S. Dresselhaus, Chem. Phys. Lett. 348 (2001) 187.
\bibitem{lozovik03} Yu.E. Lozovik, A.V. Minogin, A.M. Popov, Phys. Lett. A 313 (2003) 112.
\bibitem{lozovik03a} Yu.E. Lozovik, A.V. Minogin, A.M. Popov, JETP Lett. 77 (2003) 631.
\bibitem{wang08} X. Wang, Q. Jiang, Nanotechnology 19 (2008) 085708.
\bibitem{kang09} J.W. Kang, J.H. Lee, K.S. Kim, Y.G. Choi, Modelling Simul. Mater. Sci. Eng. 17
(2009) 025011.
\bibitem{benedict98} L.X. Benedict, N.G. Chopra, M.L. Cohen, A. Zettl, S. G. Louie, V.H. Crespi, Chem. Phys. Lett.  286 (1998) 490.
\bibitem{zacharia04} R. Zacharia, H. Ulbricht, T. Hertel, Phys. Rev. B 69 (2004) 155406.
\bibitem{girifalco56} L.A. Girifalco, R.A. Lad, J. Chem. Phys. 25 (1956) 693.
\bibitem{soule68} D. E. Soule, C.W. Nezbeda, J. Appl. Phys. 39 (1968) 5122.
\bibitem{kolmogorov00} A.N. Kolmogorov, V.H. Crespi, Phys. Rev. Lett. 85 (2000) 4727.
\bibitem{damnjanovic02} M. Damnjanovi{\'c}, T. Vukovi{\'c}, I. Milo{\v s}evi{\'c}, Eur. Phys. J. B
 25 (2002) 131.
\bibitem{vukovic03} T. Vukovi{\'c}, M. Damnjanovi{\'c}, I. Milo{\v s}evi{\'c}, Physica E 16 (2003)
256.
\bibitem{damnjanovic03} M. Damnjanovi{\'c}, E. Dobard{\v z}i{\'c}, I. Milo{\v s}evi{\'c}, T. Vukovi{\'c},
B. Nikoli{\'c}, New J. Phys. 5 (2002) 148.
\bibitem{belikov04} A.V. Belikov, A.G. Nikolaev, Yu.E. Lozovik, A.M. Popov, Chem. Phys. Lett. 385 (2004) 72.
\bibitem{bichoutskaia05} E. Bichoutskaia, A.M. Popov, A. El-Barbary, M.I. Heggie, Yu.E. Lozovik, Phys. Rev. B 71 (2005) 113403.
\bibitem{charlier93} J.-C. Charlier, J.P. Michenaud, Phys. Rev. Lett. 70 (1993) 1858.
\bibitem{kwon98} Y.K. Kwon, D. Tomanek, Phys. Rev. B 58 (1998) 16001.
\bibitem{palser99} A.H.R. Palser, Phys. Chem. Chem. Phys. 1 (1999) 4459.
\bibitem{bichoutskaia06} E. Bichoutskaia, A.M. Popov, M.I. Heggie, Yu.E. Lozovik, Phys. Rev. B 73 (2006) 045435.
\bibitem{popov09} A.M. Popov, Yu.E. Lozovik, A.S. Sobennikov, A.A. Knizhnik, JETP 108 (2009) 621.
\bibitem{kolmogorov05} A.N. Kolmogorov, V.H. Crespi, Phys. Rev. B 71 (2005) 235415.
\bibitem{xiao07} J. Xiao, B. Liu, Y. Huang, J. Zuo, K.-C. Hwang, M.-F. Yu, Nanotechnology 18 (2007) 395703.
\bibitem{class} R. Saito, M. Fujita, G. Dresselhaus, M.S. Dresselhaus, Appl. Phys. Lett. 60 (1992) 2204.
\bibitem{class1} R.A. Jishi, M.S. Dresselhaus, G. Dresselhaus, Phys. Rev. B 47 (1993) 16671.
\bibitem{lozovik04} Yu.E. Lozovik, A.M. Popov, Fullerenes, Nanotubes, Carbon Nanostructures 12 (2004) 485.
\bibitem{yan06} Q. Yan, G. Zhou, S. Hao, J. Wu, W. Duan, Appl. Phys. Lett. 88 (2006) 173107.
\bibitem{bichoutskaia06a} E. Bichoutskaia, A.M. Popov, M.I. Heggie, Yu.E. Lozovik, Fullerenes, Nanotubes
and Carbon Nanostructures 14 (2006) 131.
\bibitem{bichoutskaia07} E. Bichoutskaia, A.M. Popov, Yu.E. Lozovik, G.S. Ivanchenko, N.G. Lebedev, Phys. Lett. A 366 (2007) 480.
\bibitem{popov07} A.M. Popov, E. Bichoutskaia, Yu.E. Lozovik, A.S. Kulish, Phys. Stat. Sol. (a) 204 (2007) 1911.
\bibitem{kuznetsov07} S.S. Kuznetsov, Yu.E. Lozovik, A.M. Popov, Phys. Solid State 49 (2007) 1004.
\bibitem{bichoutskaia09a} E. Bichoutskaia, A. M. Popov, Yu. E. Lozovik, O. V. Ershova, I.V. Lebedeva, A.A. Knizhnik,
Phys. Rev. B 80 (2009) 165427.
\bibitem{kim06} H.-Y. Kim, J.D. Sofo, D. Velegal, M.W. Cole, A.A. Lucas, J. Chem. Phys. 124 (2006) 074504.
\bibitem{lozovik00} Yu. E. Lozovik, A.M. Popov, Chem. Phys. Lett. 328 (2000) 355.
\bibitem{lozovik02} Yu. E. Lozovik, A.M. Popov, Physics of the Solid State 44 (2002) 186.
\bibitem{bellarosa06} L. Bellarosa, E. Bakalis, M. Melle-Franco, F. Zerbetto, Nano Lett. 6 (2006) 1950.
\bibitem{vardanega10} D. Vardanega, F. Picaud, C. Girardet, J. Chem. Phys. 132 (2010) 124704.
\bibitem{salehinia11} I. Salehinia, S.N. Medyanik, Journal of Computational and Theoretical Nanoscience 8 (2011) 179.
\bibitem{damnjanovic99} M. Damnjanovi{\'c}, I. Milo{\v s}evi{\'c}, T. Vukovi{\'c}, R. Sredanovi{\'c}, Phys. Rev. B 60
(1999) 2728.
\bibitem{bichoutskaia09} E. Bichoutskaia, O.V. Ershova, Yu.E. Lozovik, A.M. Popov, Technical Physics Letters, 35, 666 (2009).
\bibitem{collins01} P.G. Collins, M.S. Arnold, Ph. Avouris, Science 292 (2001) 706.
\bibitem{collins01a} P.G. Collins, M. Hersam, M. Arnold, R. Martel, Ph. Avouris, Phys. Rev. Lett. 86 (2001) 3128.
\bibitem{zhang08} L. Zhang, Z. Jia, L. Huang, S. O'Brien, Z.H. Yu, J. Phys. Chem. C 112 (2008) 13893.
\end{thebibliography}
\end{document}